\documentclass[12pt,letterpaper]{article}
\pdfoutput=1

\usepackage[utf8]{inputenc}
\usepackage{amsmath}
\usepackage{amssymb}
\usepackage{amsthm}
\usepackage{xcolor}
\definecolor{nicered}{rgb}{0.7,0.1,0.1}
\definecolor{nicegreen}{rgb}{0.1,0.5,0.1}
\usepackage[colorlinks=true,citecolor= nicegreen,linkcolor=nicered]{hyperref}
\usepackage{graphicx}
\usepackage{cancel}

\usepackage{color}
\usepackage{soul}

\newcommand{\gev}[1]{{GeV} #1}

\title{Neutrino masses in the $SU(4)_{L} \otimes U(1)_{X}$ electroweak extension of the standard model}

\author{Guillermo Palacio\footnote{\texttt{galberto.palacio@udea.edu.co}}\\
Instituto de Física, Universidad de Antioquia, AA1226 \\
galberto.palacio@udea.edu.co}

\begin{document}
\maketitle

\begin{abstract}
We study the neutrino mass generation 
in the $ SU(4)_L \otimes U(1)_X $ electroweak extension
of the standard model by considering non-renormalizable
dimension five effective operators.
It is shown that there exist two topologies  
for the realizations of such an operator at the tree-level
and for one of the  three-family models is explore the neutrino phenomenology
after extending its particle content with 
an $SU(4)_L$  fermion  singlet and a scalar decuplet.
Constraints in the available parameters space of 
the model are partially discussed.
\end{abstract}

\section{Introduction}

The Standard Model (SM) of particle physics remains as one of
the most successful theories in Nature. Despite its triumph,  one of
the most direct evidences that the SM is not the final theory 
is based on the fact that neutrinos do   
oscillate{~\cite{Fukuda:1998mi,Ahmad:2002jz,Ahmad:2002ka,Ahn:2006zza}},
implying necessarily that they are massive particles. 
In the SM neutrinos are massless due to the absence of right-handed neutrinos,
that are needed to build up a Dirac mass term 
in an analogous way as is done for the charged leptons. 
In order to accommodate neutrino masses, the model must be  extended.
Among the solutions to the neutrino problem, one of the 
simplest is given by the tree level realization of the Weinberg 
operator{~\cite{Weinberg:1979sa}}, which gives rise to the
well-known type-I{~\cite{Minkowski:1977sc,Mohapatra:1979ia}}, 
type-II~\cite{Magg:1980ut,Schechter:1980gr,Cheng:1980qt,Gelmini:1980re}
and type-III~\cite{Foot:1988aq} seesaw mechanism, in where, 
an $SU(2)$ --fermion singlet, scalar triplet and fermion 
triplet-- are added respectively.
On the other hand, the SM also lacks the 
explanation for the numbers of fermion generations
in Nature. In the electroweak extension based on the 
$SU(3)_{C} \otimes SU(N)_{L} \otimes U(1)_{X}$~\cite{Pisano:1991ee,
Foot:1994ym,Riazuddin:2008yx,Gutierrez:2004sba,Ponce:2002fv}
($3$-$N$-$1$ extension for short)
gauge group, for $N \in \lbrace 3, 4 \rbrace$, the  $SU(2)_{L}$ is
enlarge to $SU(N)_{L} $. The new fermion content is 
accomodated into different fundamental
representations, $N$ or $\overline{N}$ of $SU(N)_{L}$.
From a theoretical point of view, the $3$-$N$-$1$ extension 
can account for the number fermion
generations in Nature, when the anomaly 
cancellation takes place between families and not family 
by family as in the SM{~\cite{Ponce:2002fv,Gutierrez:2004sba}}.
The electroweak $SU(4)_{L} \otimes U(1)_{X}$ also arises from  little higgs{~\cite{Lee:2014kna}} model,
provides an explanation for the charge 
quantization{~\cite{Cabarcas:2013hwa}}, allow 
electroweak unification{~\cite{Riazuddin:2008yx}} and for some kind
of models, the muon anomalus magnetic moment{~\cite{Cogollo:2014aka,Cogollo:2014tra}} is
explained within the $3$-$4$-$1$ framework.
We focus on the $3$-$4$-$1$ electroweak extension, 
which at low energies leads to
a two higgs doublet model. In this extension 
neutrinos are naturally massless, and a mechanism
for neutrino mass generation is explore through 
non-renormalizable dimension
five operator (Weinberg-like operator).
In this paper, we make a classification of the Weinberg-like
operators in a set of four three-family models.
For the so-called model $F$, we explain neutrino masses and mixing
through  the canonical seesaw mechanism and the type II-like seesaw mechanism.
For the latter case, after extending model $F$ with a scalar decuplet,
the exotic neutrinos and the lightest SM neutrinos has the same
mixing matrix and mass hierarchy.
This model has tree-level
lepton flavor violation (LFV) processes,
being $\mu \to 3e$ the most sensitive, induced by
doubly charged scalar $H_{1}^{++}$ and  controlled by its
yukawa coupling to the fermion sector $y_{\alpha \beta}$.
This article is organized as follows. 
In section~\ref{sec:Models}, the 3-4-1 electroweak extension
is reviewed, in the section~\ref{sec:Effective-Operetor} 
we classified the set of non-renormalizable effective operator
in different models of the 3-4-1 extension. A   
mechanism for neutrino mass generation in the model $F$ is explore 
through seesaw-like mechanism in the section~\ref{sec:neutrino_masses}. Finally
we summarize our main results in section~\ref{sec:conclusions}.

\section{$SU(4)_{L} \otimes U(1)_{X}$ models \label{sec:Models}}
In this section the 3-4-1 electroweak extension is 
briefly introduced.  A full phenomenological
study can be found in references{~\cite{Gutierrez:2004sba, Riazuddin:2008yx,Palcu:2009kb}}. 
We focus in the lepton sector, due that our aim is to implement higher 
dimensional effective operators
that can account for the neutrino mass generation at the tree-level.
In the electroweak  $SU(4)_{L} \otimes U(1)_{X}$, the electric charge operator is a
linear combination of the diagonal generators from the Cartan subalgebra.
\begin{eqnarray} \label{eq:elecChargeOperator}
Q =  a T_{3L} + \dfrac{b}{\sqrt{3}} T_{8L} + \dfrac{c}{\sqrt{6}} T_{15L} + X I_{4},  
\end{eqnarray}
where $a=1$ is taken in order to reproduce the SM phenomenology. The $T_{iL}$ are the generators
of $SU(4)_{L}$, normalize as $Tr(T_{i}T_{j})=\delta_{ij}/2$,  $X$ the hypercharge and $I_{4}$ the
$4 \times 4$ identity matrix. The coefficients $b$ and $c$ remains as free parameter that need
to be chosen for reach a model in particular. After demanding models that include  particles
without exotic electric charge{~\cite{Ponce:2006vw}}, two different assignments for the free parameters
are allowed. The first one, based on the selection of $b=1 \ (-1)$ and
$c=1 \ (-1)$ which gives rise to two
three-family models, Model A and Model B, and the other choice for the free parameters
is $b=1 \ (-1)$ and $c=-2 \ (2)$ that also gives rise to two three-family models, Model E
and Model F\footnote{The three-family models for the parameter assignments
  $b=-1$, $c=-1$ ($b=-1$, $c=2$) are  equivalent
  by hypercharge transformation 
to the models obtained for $b=1$, $c=1$ ($b=1$, $c=-2$). }. 

The electroweak gauge boson sector are content in the $SU(4)_{L}$ adjoint representation.
There are a total of 15 of them, which can be written as:
\begin{eqnarray} \label{eq:GaugeBosons}
\dfrac{1}{2}\lambda_{\alpha}A_{\mu}^{\alpha} = \begin{pmatrix}
D_{1\mu}^{0} & W_{\mu}^{+} & K_{\mu}^{(b+1)/2} & X_{\mu}^{(3+b+2c)/6} \\ 
W_{\mu}^{-} & D_{2\mu}^{0} & K_{1\mu}^{(b-1)/2} & V_{\mu}^{(-3+b+2c)/6} \\ 
K_{\mu}^{-(b+1)/2}  & K_{1\mu}^{-(b-1)/2} & D_{3\mu}^{0} & Y_{\mu}^{-(b-c)/3} \\ 
X_{\mu}^{-(3+b+2c)/6} & V_{\mu}^{(3-b-2c)/6} & Y_{\mu}^{(b-c)/3} & D_{4\mu}^{0}  
\end{pmatrix} \ .
\end{eqnarray}
For $b=1$ and $c=1$ in the electric charge generator, we reach two three-family
models called Model A and Model B. For the propose of this work only the lepton
and scalar sector are needed, however, for completeness also the quark sector is
displayed in table~\ref{tab:modelsAB}.
\begin{table}[t!]
\caption{Particle content for models A and B, the $\alpha= \lbrace 1,2,3 \rbrace$ are the lepton
  generation indices, $i$  run over the first two generations of quarks.
The numbers in parentheses refer to the $(SU(3)_C, SU(4)_L, U(1)_X)$ quantum numbers respectively. }
{\begin{tabular}{|c|c|} 
\hline
Model A & Model B \\
\hline
\begin{tabular}{cc} 
\\
$L_{L\alpha} =(\begin{array}{cccc}
 e^{-},  & \nu^{0}, & N^{0}, & N^{\prime 0}
\end{array})_{L\alpha} \sim (1, \overline{4}, -1/4) $,
\\
$e_{L\alpha}^{+} \sim (1,1,1)$,\\

$Q_{iL} =(\begin{array}{cccc}
 u_{i},  & d_{i}, & D_{i}, & D^{\prime}_{i}
\end{array}) \sim (3, 4, -1/12) $,
\\
$u_{iL}^{c} \sim (\overline{3}, 1, -2/3) , d_{iL}^{c} \sim (\overline{3}, 1, 1/3)  $,
\\
$D_{iL}^{c} \sim (\overline{3}, 1, 1/3) ,D_{iL}^{\prime c} \sim (\overline{3}, 1, 1/3)  $,
\\
$Q_{3L} =(\begin{array}{cccc}
 d_{3},  & u_{3}, & U_{3}, & U^{\prime}_{3}
\end{array}) \sim (3, \overline{4}, 5/12) $,
\\
$u_{3L}^{c} \sim (\overline{3}, 1, -2/3) , d_{3L}^{c} \sim (\overline{3}, 1, 1/3)  $, 
\\
$U_{3L}^{c} \sim (\overline{3}, 1, -2/3) ,U_{3L}^{\prime c} \sim (\overline{3}, 1, -2/3)  $, \\
\end{tabular}

&

\begin{tabular}{cc} 
\\
$L_{L\alpha} =(\begin{array}{cccc}
 \nu^{0}, e^{-} & , & E^{-}, & E^{\prime -}
\end{array})_{L\alpha} \sim (1, 4, -3/4) $,\\ 

$e_{L\alpha}^{+} \sim (1,1,1)$, $E_{L\alpha}^{+} \sim (1,1,1)$, $E^{\prime +}_{L\alpha} \sim (1,1,1)$,\\

$Q_{iL} =(\begin{array}{cccc}
 d_{i},  & u_{i}, & U_{i}, & U^{\prime}_{i}
\end{array}) \sim (3, \overline{4}, 5/12) $,\\ 

$u_{iL}^{c} \sim (\overline{3}, 1, -2/3) , d_{iL}^{c} \sim (\overline{3}, 1, 1/3)  $, \\

$U_{iL}^{c} \sim (\overline{3}, 1, -2/3) ,U_{iL}^{\prime c} \sim (\overline{3}, 1, -2/3)  $, \\

$Q_{3L} =(\begin{array}{cccc}
 u_{3},  & d_{3}, & D_{3}, & D^{\prime}_{3}
\end{array}) \sim (3, 4, -1/12) $,\\ 

$u_{3L}^{c} \sim (\overline{3}, 1, -2/3) , d_{3L}^{c} \sim (\overline{3}, 1, 1/3)  $, \\

$D_{3L}^{c} \sim (\overline{3}, 1, 1/3) ,U_{3L}^{\prime c} \sim (\overline{3}, 1, 1/3)  $, \\

\end{tabular}
\\
\hline
\end{tabular} \label{tab:modelsAB}}
\end{table}
The scalar sector for this set of models is given by:
\begin{eqnarray} \label{eq:00003}
\left\langle \Phi _{1}^{T}\right\rangle & = &\left\langle \left( \phi
_{1}^{0},\phi _{1}^{-},\phi _{1}^{\prime -},\phi _{1}^{\prime \prime -}\right)
\right\rangle =\left( v,0,0,0\right) \sim \left( 1,4 ,-3/4\right) , \nonumber \\
\left\langle \Phi _{2}^{T}\right\rangle & =& \left\langle \left( \phi
_{2}^{+},\phi _{2}^{0},\phi _{2}^{\prime 0},\phi _{2}^{\prime \prime 0}\right)
\right\rangle =\left( 0,v^{\prime},0,0\right) \sim \left( 1,4,1/4
\right) ,\nonumber \\ 
\left\langle \Phi _{3}^{T}\right\rangle & =&\left\langle \left( \phi
_{3}^{+},\phi _{3}^{0},\phi _{3}^{\prime 0},\phi _{3}^{\prime \prime 0}\right)
\right\rangle =\left( 0,0,V,0\right) \sim \left( 1,4 ,1/4\right), \nonumber \\ \label{vev}
\left\langle \Phi _{4}^{T}\right\rangle & =&\left\langle \left( \phi
_{4}^{+},\phi _{4}^{0},\phi _{4}^{\prime 0},\phi _{4}^{\prime \prime 0}\right)
\right\rangle =\left( 0,0,0,V^{\prime}\right) \sim \left( 1,4,1/4
\right).
\end{eqnarray}
For $b=1$ and $c=-2$ in the electric charge generator we reach two three-family
models called Model E and Model F, which are displayed in table~\ref{tab:modelsEF}.
\begin{table}[b!]
\caption{Particle content for models E and F, the $\alpha= \lbrace 1,2,3 \rbrace$ are the
  lepton generation indices, $i$  run over the first two generations of quarks.
The numbers in parentheses refer to the $(SU(3)_C, SU(4)_L, U(1)_X)$ quantum numbers respectively. }
{\begin{tabular}{|c|c|} 
\hline
Model E & Model F \\
\hline
\begin{tabular}{cc} 
\\
 $L_{L\alpha} =(\begin{array}{cccc}
e^{-}, & \nu^{0},  & N^{0}, & E^{-}
\end{array})_{L\alpha} \sim (1, \overline{4}, -1/2) $, \\ 

$e_{L\alpha}^{+} \sim (1,1,1)$, $E_{L\alpha}^{+} \sim (1,1,1)$,\\

$Q_{iL} =(\begin{array}{cccc}
 u_{i},  & d_{i}, & D_{i}, & U_{i}
\end{array}) \sim (3, 4, 1/6) $,\\ 

$u_{iL}^{c} \sim (\overline{3}, 1, -2/3) , d_{iL}^{c} \sim (\overline{3}, 1, 1/3)  $,\\

$U_{iL}^{c} \sim (\overline{3}, 1, -2/3) ,D_{iL}^{c} \sim (\overline{3}, 1, 1/3)  $, \\

$Q_{3L} =(\begin{array}{cccc}
 d_{3},  & u_{3}, & U_{3}, & D_{3}
\end{array}) \sim (3, \overline{4}, -1/12) $,\\ 

$u_{3L}^{c} \sim (\overline{3}, 1, -2/3) , d_{3L}^{c} \sim (\overline{3}, 1, 1/3)  $, \\

$U_{3L}^{c} \sim (\overline{3}, 1, -2/3) ,D_{3L}^{c} \sim (\overline{3}, 1, 1/3)  $, \\

\end{tabular}

&

\begin{tabular}{cc} 
\\
 $L_{L\alpha} =(\begin{array}{cccc}
 \nu^{0}, &  e^{-},  & E^{-}, & N^{0}
\end{array})_{L\alpha} \sim (1, 4, -1/2) $, \\

$e_{L\alpha}^{+} \sim (1,1,1)$, $E_{L\alpha}^{+} \sim (1,1,1)$,\\

$Q_{iL} =(\begin{array}{cccc}
 d_{i},  & u_{i}, & U_{i}, & D_{i}
\end{array}) \sim (3, \overline{4}, 1/6) $,\\ 

$u_{iL}^{c} \sim (\overline{3}, 1, -2/3) , d_{iL}^{c} \sim (\overline{3}, 1, 1/3)  $,\\

$U_{iL}^{c} \sim (\overline{3}, 1, -2/3) ,D_{iL}^{c} \sim (\overline{3}, 1, 1/3)  $, \\

$Q_{3L} =(\begin{array}{cccc}
 u_{3},  & d_{3}, & D_{3}, & U_{3}
\end{array}) \sim (3, 4, 1/6) $,\\ 

$u_{3L}^{c} \sim (\overline{3}, 1, -2/3) , d_{3L}^{c} \sim (\overline{3}, 1, 1/3)  $, \\

$D_{3L}^{c} \sim (\overline{3}, 1, 1/3) ,U_{3L}^{c} \sim (\overline{3}, 1, -2/3)  $, \\

\end{tabular}
\\
\hline
\end{tabular} \label{tab:modelsEF}}
\end{table}
The scalar sector for this set of models is given by:
\begin{eqnarray} \label{eq:00004}
\left\langle \Phi _{1}^{T}\right\rangle & = &\left\langle \left( \phi
_{1}^{0},\phi _{1}^{+},\phi _{1}^{\prime +},\phi _{1}^{\prime 0}\right)
\right\rangle =\left( v,0,0,0\right) \sim \left( 1,\overline{4} ,1/2\right) ,\nonumber \\ 
\left\langle \Phi _{2}^{T}\right\rangle & = & \left\langle \left( \phi
_{2}^{-},\phi _{2}^{0},\phi _{2}^{\prime 0},\phi _{2}^{\prime -}\right)
\right\rangle =\left( 0,v^{\prime},0,0\right) \sim \left( 1,\overline{4},-1/2
\right) , \nonumber \\ 
\left\langle \Phi _{3}^{T}\right\rangle & = &\left\langle \left( \phi
_{3}^{-},\phi _{3}^{0},\phi _{3}^{\prime 0},\phi _{3}^{\prime -}\right)
\right\rangle =\left( 0,0,V,0\right) \sim \left( 1,\overline{4},-1/2\right), \nonumber \\
\left\langle \Phi _{4}^{T}\right\rangle & = &\left\langle \left( \phi
_{4}^{0},\phi _{4}^{+},\phi _{4}^{\prime +},\phi _{4}^{\prime 0}\right)
\right\rangle =\left( 0,0,0,V^{\prime}\right) \sim \left( 1,\overline{4},1/2
\right).
\end{eqnarray}
The pattern  of the electroweak symmetry breaking (EWSB) goes as follows
\begin{eqnarray}{\label{breaking_pattern}}
  SU(4)_{L} \otimes  U(1)_X \xrightarrow{V^{\prime}} SU(3)_{L} \otimes  U(1)_{X^{\prime}} 
\xrightarrow{V}   SU(2)_{L} \otimes  U(1)_{Y} \xrightarrow{v,v^{\prime}}  U(1)_{\rm{Q}} ,  
\end{eqnarray}
where $V^{\prime} \sim V \gg v^{\prime} \sim v $,
and ${v^{\prime}}^{2} +v^{2} =v_{\rm{SM}}^{2} \equiv (246 \ \rm{GeV} )^{2} $.

\section{Dimension 5  effective operator  \label{sec:Effective-Operetor}}
Neutrinos may acquire masses after the introduction of non-renormalizable 
dimension-five operators defined as:
\begin{eqnarray} \label{eq:Dimesion5Operator}
  {\mathcal{L}}_{5} =  \dfrac{\mathcal{O}_{5} }{\Lambda}, \ \ \
  \mathcal{O}_{5} =\lbrace \overline{L_{L\alpha}^{c}}\Phi_{i}^{\star}\Phi_{j}^{\dagger}L_{L\beta},
  \overline{L_{L\alpha}^{c}}\Phi_{i}\Phi_{j}^{\star \dagger}L_{L\beta}\rbrace,
\end{eqnarray}
being $\alpha$ and $\beta$ lepton generation indices and $i$, $j$ index in the number of
scalar 4-plets. $\Lambda$ represent the cutoff scale where new physics is expected.
The operator given in Eq.~(\ref{eq:Dimesion5Operator}) is the generalization
of the Weinberg operator~\cite{Weinberg:1979sa} for $SU(4)_{L} \otimes U(1)_{X}$.
Depending on the way as the fields transforms under $SU(4)_{L} \otimes U(1)_{X}$,
different tree-level realizations of the operator are allowed.


\begin{table}[h!]
\caption{Scenarios for the operator defined in
  Eq.~(\ref{eq:Dimesion5Operator}): In the left part,
  the ($4(\overline{4}),X_{L(\Phi)} )$ notation represents the way as the
  fields (either $L_{L\alpha}$ or $\Phi_{i}$) transforms under
  $SU(4)_{L} \otimes U(1)_{X}$. The effective operator is allowed if it is gauge invariant.}
{\begin{tabular}{|c|c|c|c|} 
\hline
$L_{L \alpha}$ & $\Phi_k$ & $\mathcal{O}_{5}^{I} = \overline{L_{L\alpha}^{c}}\Phi_{i}^{\star}\Phi_{j}^{\dagger}L_{L\beta}$
 & $\mathcal{O}_{5}^{II}=\overline{L_{L\alpha}^{c}}\Phi_{i}\Phi_{j}^{\star \dagger}L_{L\beta}$  
\\
\hline
\hline

$(4,X_{L})$ & $(4,X_{\Phi})$ & 
$4\otimes\overline{4}\otimes\overline{4}\otimes4 \supset 1 $, \ $2X_{L} - 2X_{\Phi} = 0  $ 
 & $4\otimes4\otimes4\otimes4 \supset  1 $, \ $2X_{L} + 2X_{\Phi} = 0 $  \\
\hline 
$(4,X_{L})$ & $(\overline{4},X_{\Phi})$ & $4\otimes4\otimes4\otimes4 \supset  1 $, \ $2X_{L} - 2X_{\Phi} = 0  $  & $4\otimes\overline{4}\otimes\overline{4}\otimes4  \supset 1 $, \ $2X_{L} + 2X_{\Phi} = 0 $  \\ 
\hline 
$(\overline{4},X_{L})$ & $(4,X_{\Phi})$  & $\overline{4}\otimes\overline{4}\otimes\overline{4}\otimes\overline{4} \supset  1  $, \ $2X_{L} - 2X_{\Phi} = 0  $  & $\overline{4}\otimes4\otimes4\otimes\overline{4} \supset  1 $,
\ $2X_{L} + 2X_{\Phi} = 0 $  \\ 
\hline 
$(\overline{4},X_{L})$ & $(\overline{4},X_{\Phi})$  & $\overline{4}\otimes 4 \otimes 4\otimes\overline{4} \supset 1 $, \ $2X_{L} - 2X_{\Phi} = 0  $ & $\overline{4}\otimes\overline{4}\otimes\overline{4}\otimes\overline{4} \supset 1 $, \ $2X_{L} + 2X_{\Phi} = 0 $  \\ 
\hline 

\end{tabular} \label{tab:operators_set}}
\end{table}

For any set of fields $(\Phi, L_{L})$, transforming
in a general way under $SU(4)_{L} \otimes U(1)_{X}$, different theoretical realizations of
the operators are displayed in table \ref{tab:operators_set}.
In order to allow such an operators into an effective lagrangian, we must garanteed that the
product of the irreducible representations contain the $SU(4)$ singlet and be hyperchargeless.
Since for $SU(N)$,  $N \otimes N = [(N^{2}+N)/2]_{\rm{S}} +  [(N^{2}-N)/2]_{\rm{A}} $ and
$N \otimes N^{\star} = [N^{2}-1]_{\rm{Adjoint}} +  [1] $, there are only two possible
main topologies for the tree-level realization of the Weinberg operator.
From Eq.~(\ref{eq:Dimesion5Operator}), if the intermediate particle is a scalar,
it can transform as $10_{\rm{S}}$ and $15_{\rm{Adjoint}}$\footnote{The scalar singlet
does not gives rise to neutrino masses.} under $SU(4)_{L}$, on the other hand if it
is a fermion, it can transform as $1_A$,  and $15_{\rm{Adjoint}}$ under
$SU(4)_{L}$\footnote{
The fermion sextet is also a possible
realization of the Weinger operator, however is not allowed
because after their introduction it does not give rise to neutrino masses,
instead  is an additional term that contribute to the masses of the charges
leptons.}. 
\begin{figure}
\centering
\includegraphics[scale=0.7]{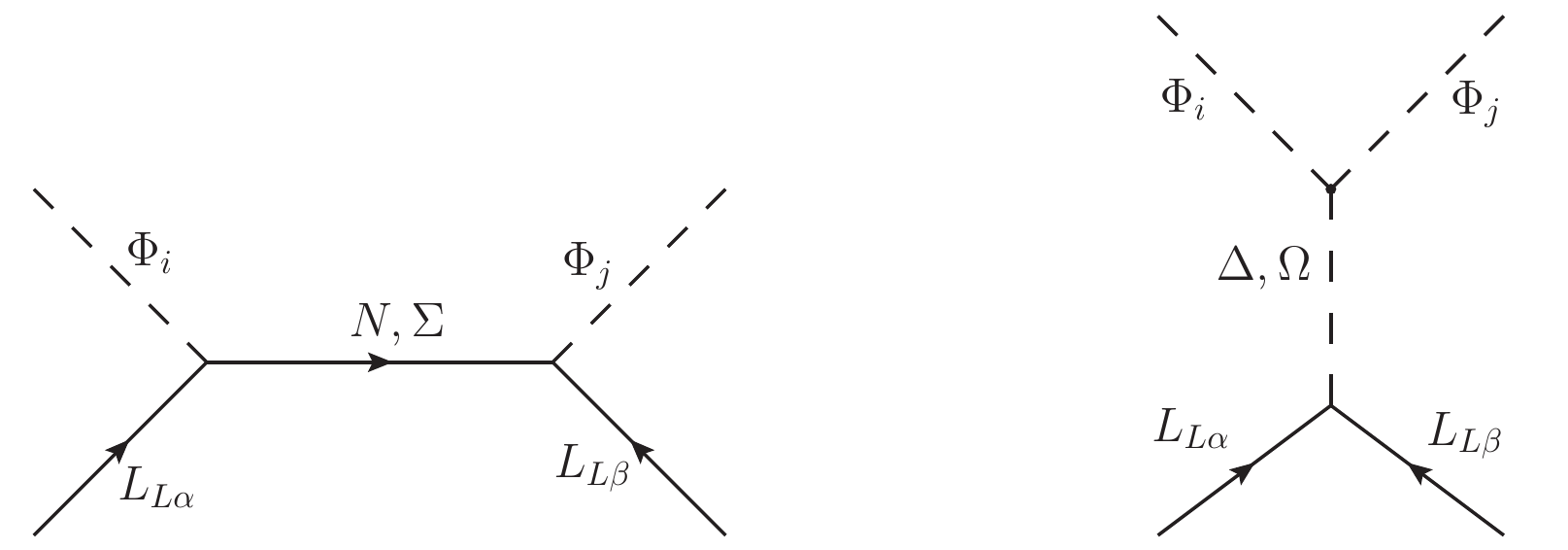}
\caption{Topologies of the Weinberg-like
effective operator. On the left hand side  the intermediate particle could be an $SU(4)$
fermion singlet $N_{R} \sim (1,0)$,  and a fermion 15-plet $\Sigma \sim (15,0)$. 
On the right hand side  the intermediate particles could be an
$SU(4)$ scalar decuplet $\Delta \sim (10,X_{\Delta})$ and a scalar 15-plet $\Omega \sim (15,X_{\Omega})$. }
{\label{figgg1:tree_level}}
\end{figure}
In Figure~\ref{figgg1:tree_level}  are displayed all the possible
tree level realization of the effective Weinberg operator in the $SU(4)_L \otimes U(1)_X$ 
electroweak extension. The theory
reduces to a canonical seesaw, a type II-like seesaw, 
and a type III-like seesaw in where, for $SU(4)_L$ a -fermion singlet,  scalar
decuplet and fermion 15-plet - are included respectively.

To our knowledge the 3-4-1 extension with a fermion singlet (canonical seesaw mechanism)
has been implemented{~\cite{Palcu:2009uk}},
as well as with a scalar decuplet{~\cite{Foot:1994ym, Long:2016lmj}}, but the fermion
15-plet has not been proposed in the literature yet.
Those new particles in case of be added  should have hypercharge
values that does not spoil the anomaly free structure of the model.
That is why any new fermion content should
have zero hypercharge  or be a vector-like particle under $SU(4)_{L}$. 
In the next subsections, we  display the set of 
effective Weinberg-like operators that can be built in the four models presented in section~\ref{sec:Models}.
\subsection{Model A}
In this model there are a total of 9 operators,  which are given by:
\begin{eqnarray}\label{eq:TotalSetOperatorModelA} \nonumber
  {\mathcal{O}}_{5} &=& \Big\lbrace \overline{L_{L\alpha}^{c}}
  \Phi_{2} \Phi_{2}^{\star \dagger} L_{L\beta},
 \ \ \overline{L_{L\alpha}^{c}} \Phi_{2} \Phi_{3}^{\star\dagger}L_{L\beta},
\ \ \overline{L_{L\alpha}^{c}} \Phi_{3} \Phi_{2}^{\star\dagger}L_{L\beta},
 \\ \nonumber && \overline{L_{L\alpha}^{c}} \Phi_{2} \Phi_{4}^{\star\dagger}L_{L\beta},
  \ \ \overline{L_{L\alpha}^{c}} \Phi_{4} \Phi_{2}^{\star\dagger}L_{L\beta},
 \ \ \overline{L_{L\alpha}^{c}} \Phi_{3} \Phi_{3}^{\star \dagger}L_{L\beta},
  \\ && \overline{L_{L\alpha}^{c}} \Phi_{3}\Phi_{4}^{\star\dagger}L_{L\beta},
 \ \ \overline{L_{L\alpha}^{c}} \Phi_{4} \Phi_{3}^{\star \dagger}L_{L\beta},
  \ \ \overline{L_ {L\alpha}^{c}} \Phi_{4} \Phi_{4}^{\star \dagger}L_{L\beta} \Big\rbrace \ .
\end{eqnarray}
For this model we have:
\begin{enumerate}
\item $\Phi_{k}\Phi_{k}^{\star \dagger} \Rightarrow 4 \otimes 4 = 6_{\rm{A}} \oplus 10_{\rm{S}}  $,
  therefore a $10_{\rm{S}}$  scalar is allowed as the intermediate particle,
  the $6_{\rm{A}}$ is not allowed because of its statistic. 
\item $\Phi_{k}^{\star \dagger}L_{L\beta} \Rightarrow 4 \otimes \overline{4} = 1 \oplus 15_{\rm{Adjoint}} $,
  then either a fermion singlet or a fermion 15-plet are allowed as intermediate particles.  
\end{enumerate}

The operators defined in Eq.~(\ref{eq:TotalSetOperatorModelA})  have two topologies
at the tree-level, one in which the intermediate particle
is a fermion, either singlet $N_{R} \sim (1,1,0)$  or 15-plet $\Sigma \sim (1,15,0)$, and
the other one in which the intermediate particle is a scalar decuplet $\Delta \sim (1,10,1/2)$.
In order to fit all the experimental neutrino oscillation parameters,
at least three right-handed neutrinos ( three fermion 15-plet ) per
lepton generation  or an scalar decuplet must be added.

\subsection{Model B}

For this model the operator is unique and is given by:
\begin{eqnarray}\label{eq:00009}
{\mathcal{O}}_{5} = \Big \lbrace \overline{L_{L\alpha}^{c}} \Phi_{1}^{\star} \Phi_{1}^{\dagger}L_{L\beta} \Big \rbrace \ .
\end{eqnarray}
\begin{enumerate}
\item $\Phi_{k}^{\star}\Phi_{k}^{\dagger} \Rightarrow \overline{4} \otimes \overline{4} =  6_{\rm{A}} \oplus
  \overline{10_{\rm{S}}}$, therefore a $\overline{10_{\rm{S}}}$
  scalar is allowed as the intermediate particle,
  the $6_{A}$ is forbidden due to its statistic. 
\item $\Phi_{k}^{\dagger}L_{L\beta} \Rightarrow \overline{4} \otimes 4 = 1 \oplus 15_{\rm{Adjoint}} $, then either a
  fermion singlet or a fermion 15-plet are allowed as intermediate particles.  
\end{enumerate}

The operators given in Eq.~(\ref{eq:00009}) has two topologies at tree level,
one in which the intermediate particle is a fermion, either
singlet $N_{R} \sim (1,1,0)$ or  15-plet $\Sigma \sim (1,15,0)$, and
the other one in which the intermediate particle is a scalar decuplet $\Delta \sim (1,10,3/2)$.
Again, to fit all the experimental neutrino oscillation parameters, at least one
right-handed neutrino (15-plet fermion) per lepton generation or an scalar decuplet must be included.

\subsection{Model E}

For this model there are 4  operators, which are given by:
\begin{eqnarray}\label{eq:00010} 
{\mathcal{O}}_{5} &=& \Big\lbrace \overline{L_{L\alpha}^{c}} \Phi_{2}^{\star} \Phi_{2}^{\dagger}L_{L\beta}, \ \  \overline{L_{L\alpha}^{c}} \Phi_{2}^{\star} \Phi_{3}^{\dagger}L_{L\beta}, \ \ \overline{L_{L\alpha}^{c}} \Phi_{3}^{\star} \Phi_{2}^{\dagger}L_{L\beta}, \ \ \overline{L_{L\alpha}^{c}} \Phi_{3}^{\star} \Phi_{3}^{\dagger}L_{L\beta}  \Big\rbrace \ .
\end{eqnarray}
\begin{enumerate}
\item $\Phi_{k}^{\star}\Phi_{k}^{\dagger} \Rightarrow 4 \otimes 4 = 6_{\rm{A}} \oplus
  10_{\rm{S}}$, then a $10_{\rm{S}}$
   scalar is allowed as the intermediate particle,
  the $6_{A}$ is not allowed because its statistic. 
\item $\Phi_{k}^{\dagger}L_{L\beta} \Rightarrow 4 \otimes \overline{4} = 1 \oplus 15_{\rm{Adjoint}} $,
  then either a fermion singlet or a fermion 15-plet are allowed as intermediate particles.  
\end{enumerate}

Again, each of the previous operators have two topologies
at tree level, one in which the intermediate particle is a
fermion either singlet $N_{R} \sim (1,1,0)$ or  15-plet $\Sigma \sim (1,15,0)$, and
the other one in which the intermediate particle is a scalar decuplet $\Delta \sim (1,10,1)$.
In order to fit all the experimental neutrino oscillation parameters,
at least two right-handed neutrinos ( two fermion 15-plet ) per
lepton generation  or an scalar decuplet must be added.

\subsection{Model F}

For this model there are 4  operators, which are given by:
\begin{eqnarray}\label{eq:00012} 
{\mathcal{O}}_{5} &=& \Big\lbrace \overline{L_{L\alpha}^{c}} \Phi_{1} \Phi_{1}^{\star \dagger}L_{L\beta}, \ \  \overline{L_{L\alpha}^{c}} \Phi_{1} \Phi_{4}^{\star \dagger}L_{L\beta}, \ \ \overline{L_{L\alpha}^{c}} \Phi_{4} \Phi_{1}^{\star \dagger}L_{L\beta}, \ \ \overline{L_{L\alpha}^{c}} \Phi_{4} \Phi_{4}^{\star \dagger}L_{L\beta}  \Big\rbrace \ .
\end{eqnarray}
\begin{enumerate}
\item $\Phi_{k}\Phi_{k}^{\star \dagger} \Rightarrow \overline{4} \otimes \overline{4} = 6_{\rm{A}} \oplus
  \overline{10_{\rm{S}}} $, then a $\overline{10_{\rm{S}}}$
   scalar is allowed as the intermediate particle,
  the $6_{\rm{A}}$ is not allowed because of its statistic. 
\item $\Phi_{k}^{\star \dagger}L_{L\beta} \Rightarrow \overline{4} \otimes 4 = 1 \oplus 15_{\rm{Adjoint}} $, then either a
  fermion singlet or a fermion 15-plet are allowed as intermediate particles.  
\end{enumerate}

The operators given in Eq.~(\ref{eq:00012}) have two topologies
at tree level, one in which the intermediate particle
is a fermion either singlet $N_{R} \sim (1,1,0)$
or 15-plet   $\Sigma \sim (1,15,0)$, and the other
one in which the intermediate particle is a scalar
decuplet $\Delta \sim (1,10,1)$.
Neutrino oscillation parameters are explained after the model is
 extended with two right-handed neutrinos (or two fermion 15-plets) per lepton generation or 
 a scalar decuplet.

To address neutrino masses and mixing, models 
   with fermion singlets{~\cite{Palcu:2009uk,Palcu:2015ica}} as well
   as with scalar decuplets has been constructed{~\cite{Long:2016lmj}}.
   In particular in Ref.{~\cite{Palcu:2011tw}} not new particles were introduced,
   instead the $10_S$ scalar representation was build using the 
fundamental representation of the scalar fields content in $SU(4)_L$.
Scalar decuplets also has been used to provide
masses for the charged leptons in $3$-$4$-$1$ models
{~\cite{Pisano:1994tf}}.  In the next section we study the neutrino mass generation
and mixing in the model $F$, extending with a fermion singlets,
and a scalar decuplet.  
\section{Neutrino masses in  Model F}\label{sec:neutrino_masses}
In order to explain neutrino masses and mixing in the $3$-$4$-$1$ electroweak extension,
  we explore the tree-level realization of the Weinberg-like
  operator in the  model $F$ introduced\footnote{The same
    can be done for all the models, following the general classification given in
    chapter~\ref{sec:Effective-Operetor}.} in table{~\ref{tab:modelsEF}}. 
\subsection{Canonical Seesaw Mechanism }
The model $F$ is extended with two right-handed neutrinos
$N_{{\rm{1R}}i} \sim (1,1,0)$ and $N_{{\rm{2R}}i} \sim (1,1,0)$, being $i$ the generation index. At
least three generations of $\lbrace N_{\rm{{1Ri}}}, N_{{\rm{2Ri}}} \rbrace $  are needed
in order explain the neutrino masses.
The most general Yukawa lagrangian for the neutral lepton sector,
including the new fields reads:
\begin{eqnarray}\label{eq:00014} 
  -\mathcal{L}_{\rm{yuk}} &=& \bigg [  \lambda_{1}^{\alpha i} \overline{L_{L\alpha}}\Phi_{1}N_{1Ri} +
  \lambda_{2}^{\alpha j} \overline{L_{L\alpha}}\Phi_{1}N_{2Rj} +
  \lambda_{3}^{\alpha i} \overline{L_{L\alpha}}\Phi_{4}N_{1Ri} \nonumber\\ 
  &+& \lambda_{4}^{\alpha j} \overline{L_{L\alpha}}\Phi_{4}N_{2Rj} +  h.c \bigg ]
  + \dfrac{1}{2}M_{1}\overline{N_{1Ri}^{C}}N_{1Ri} + \dfrac{1}{2}M_{2}\overline{N_{2Rj}^{C}}N_{2Rj}
  \nonumber \\ &+&
  \bigg [ \mu \overline{N_{1Ri}^{C}}N_{2Rj} + h.c \bigg ] \ ,
\end{eqnarray}
where $\lambda_{l}^{\alpha i}$; for $ l \in  \lbrace 1,2,3,4 \rbrace
$ and $i \in \lbrace 1, 2 \rbrace$, are $ 3 \times k$  Yukawa matrix  entries; $k$, 
the number of right-handed neutrinos per lepton generation, $M_1$ and $M_2$
are $3 \times 3$ Majorana mass matrices for the right-handed neutrinos and are assumed to be
diagonal without loss of generality. $\mu$ is a mixing term, that in general
is allowed by the gauge symmetry. After the electroweak symmetry breaking (EWSB),
Eq.~(\ref{eq:00014}) becomes:
\begin{eqnarray}\label{eq:00015} 
  -\mathcal{L}_{\rm{yuk}} &=& \begin{pmatrix}
  \overline{\nu_{L \alpha}} & \overline{N_{L \alpha}} & \overline{N_{1R i}^{C}} & \overline{N_{2R i}^{C}}
  \end{pmatrix}  \mathcal{M} \begin{pmatrix}
  \nu_{L \alpha} \\ 
  N_{L \alpha} \\ 
  N_{1R i} \\ 
  N_{2R i}
  \end{pmatrix} \ , 
\end{eqnarray}
with:
\begin{eqnarray}\label{eq:canseesaw_massmatrix}
\mathcal{M}=\begin{pmatrix}
0 & 0 & v\lambda_{1} & v\lambda_{2} \\ 
0 & 0 & V^{\prime}\lambda_{3} & V^{\prime}\lambda_{4} \\ 
v\lambda_{1}^{\dagger} & V^{\prime}\lambda_{3}^{\dagger} & M_1 & \mu \\ 
v\lambda_{2}^{\dagger} & V^{\prime}\lambda_{4}^{\dagger} & \mu & M_2
\end{pmatrix} \equiv \begin{pmatrix}
0 & 0 & {m_{1D}} & m_{2D} \\ 
0 & 0 & {m_{3D}} & m_{4D} \\ 
m_{1D}^{\dagger} & m_{3D}^{\dagger} & {M_1} & {\mu} \\ 
m_{2D}^{\dagger} & m_{4D}^{\dagger} & {\mu} & {M_2}
\end{pmatrix} \equiv \begin{pmatrix}
0_{6 \times 6} & M_{D} \\ 
M_{D}^{\dagger} & M_{R}
\end{pmatrix} \ .    
\end{eqnarray}
The mass matrix given in Eq.~(\ref{eq:canseesaw_massmatrix}) can not be
diagonalized exactly. However for simplicity and illustrative purposes
we set all element of matrix $\mu$ to be zero.
In this model, the smallness of active neutrinos is due to the heavyness of
the right-handed neutrinos as happens in the SM with the
type I seesaw mechanism. In the limit
$\lbrace M_{1} ,M_{2}\rbrace >> \lbrace  m_{1D},m_{2D},m_{3D},m_{4D}  \rbrace $,
the  mass matrix in Eq.~(\ref{eq:canseesaw_massmatrix})  can be diagonalized
by blocks in an approximately way, and the masses for the lightest 
and heaviest neutrinos takes the form:
\begin{eqnarray}\label{eq:neutrino_masses_expressions}
\mathcal{M}^{\rm{light}} &=& - M_{R}^{-1}M_{D}M_{D}^{\dagger} 
 + \mathcal{O}(M_{R}^{-2})
\approx - \begin{pmatrix}
\alpha &
\beta \\ 
\gamma &
 \delta
\end{pmatrix} \ ,\\
\mathcal{M}^{\rm{heavy}} &=& M_{R} +  \mathcal{O}(M_{R}^{-1}) \approx \begin{pmatrix}
M_{1} & 0 \\ 
0 & M_{2}
\end{pmatrix} \ ,
\end{eqnarray}
where
\begin{eqnarray}
\alpha  &=& M_{1}^{-1} [m_{1D}m_{1D}^{\dagger} + m_{2D}m_{2D}^{\dagger} ] \ , \nonumber \\
\beta   &=& M_{1}^{-1} [m_{1D}m_{3D}^{\dagger} + m_{2D}m_{4D}^{\dagger} ] \ , \nonumber \\
\gamma  &=& M_{2}^{-1} [m_{3D}m_{1D}^{\dagger} + m_{4D}m_{2D}^{\dagger} ] \equiv M_{2}^{-1}\beta^{\dagger} M_{1} \ , \nonumber \\
\delta  &=& M_{2}^{-1} [m_{3D}m_{3D}^{\dagger} + m_{4D}m_{4D}^{\dagger} ] \ .
\end{eqnarray}

From Eq.~(\ref{eq:neutrino_masses_expressions}), the lightest neutrino
spectrum in the physical basis is obtained as:
\begin{eqnarray}{\label{eq:neutrino_spectrum}}
\mathcal{M}_{\rm{diag}}^{\rm{light}}=U^{\dagger} \mathcal{M}^{\rm{light}}U \ ,
\end{eqnarray}

being $U$ a $6 \times 6$ matrix which mixed{~\cite{Escrihuela:2015jaa}} the lightest neutrinos

\begin{eqnarray}\label{eq:mixing_full_66}
U^{^{6 \times 6}} = \begin{pmatrix}
N^{^{3 \times 3}} & S^{^{3 \times 3}} \\ 
T^{^{3 \times 3}} & V^{^{3 \times 3}}
\end{pmatrix} .
\end{eqnarray} 
  From the experimental side, oscillations between the three   
  active SM neutrinos and exotic neutrinos have not
  yet being observed{~\cite{Aguilar-Arevalo:2013pmq}},
  implying that  new neutral leptons, if they exist, must
  be heavy, $m_{N_{L}} > 1$~eV. As a consequence,  the mixing matrices
  $S^{^{3 \times 3}}$ and $T^{^{3 \times 3}}$  in Eq.~(\ref{eq:mixing_full_66}) will be suppressed.
  As pointed out{~\cite{Escrihuela:2015jaa}}, the current experimental limits
  on neutrinos oscillation experiments are not able to put stringent constraints
  in any of the new physics (NP) parameters given inside Eq.~(\ref{eq:mixing_full_66});
  however, a future generation of neutrino experiment will open the window for
  the exploration{~\cite{Wurm:2011zn}}.
  The lepton flavor violation (LFV) processes such as $\mu \to e \gamma$ can take place
  in this model at one loop level, however a full
  study on LFV is beyond scope of this paper.  
The lightest active SM neutrinos  
acquire masses through the canonical seesaw mechanism, as happens
for the SM. Based on the above observations, the mixing matrix in 
Eq.~(\ref{eq:mixing_full_66}) is approximately diagonal\footnote{There are
not mixing between the sterile neutrinos and the SM ones.}, and the 
masses for the lightest SM neutrinos takes the form: 
\begin{eqnarray}\label{eq:Final_neutrino_masses_SM}
\mathcal{M}_{\widehat{\nu}_L}^{\rm{diag}} & \approx & N^{\dagger} \mathcal{M}_{\nu_L}  N \ , \nonumber \\
\mathcal{M}_{\widehat{\nu}_L}^{\rm{diag}} & \approx & U_{\rm{PMNS}}^{\dagger}   M_{1}^{-1} [m_{1D}m_{1D}^{\dagger}]
  U_{\rm{PMNS}}~, 
\end{eqnarray}
with $U_{\rm{PMNS}}$, the Pontecorvo-Maki-Nakagawa-Sakata
  mixing matrix{~\cite{Maki:1962mu}} and
  $\mathcal{M}_{\widehat{\nu}_L}^{\rm{diag}}={\rm{diag}}(m_{\nu1},
  m_{\nu2}, m_{\nu3} )$.
 The masses for the lightest sterile neutrinos reads,
\begin{eqnarray}\label{eq:Final_neutrino_masses_Sterile}
 \mathcal{M}_{\widehat{N}_L}^{\rm{diag}} &\approx& V^{\dagger} \mathcal{M}_{N_L}  V \ ,  \nonumber \\ 
 \mathcal{M}_{\widehat{N}_L}^{\rm{diag}} & \approx & V^{\dagger} M_{2}^{-1} [ m_{4D}m_{4D}^{\dagger} ]V^{\dagger} \ ,
\end{eqnarray} 
with $\mathcal{M}_{\widehat{N}_L}^{\rm{diag}}={\rm{diag}}(m_{N1},
  m_{N2}, m_{N3} )$.
The Eq.~(\ref{eq:Final_neutrino_masses_SM}) and Eq.~(\ref{eq:Final_neutrino_masses_Sterile})
 were obtained after demanding $\lambda_1 \gg  \lambda_2$ and
 $\lambda_4 \gg  \lambda_3$. 
Under these assumptions the two neutrino sectors are uncorrelated.
The masses for the SM neutrinos are fully determined by $M_1$, $\lambda_1$ and $U_{\rm{PMNS}}$.
\subsection{Type II-like Seesaw Mechanism}
The model F displayed in table{~\ref{tab:modelsEF}} is extended with a
scalar decuplet  $\Delta \sim (1,10,1)$. The most general lagrangian 
for the neutral
leptons is given by:
\begin{eqnarray}{\label{eq:Type_II_lag_mass}}
-\mathcal{L}_{\rm{yuk}} = y_{\alpha \beta} \overline{\widetilde{L_{L \alpha}^{C}}} \Delta L_{L \beta} + h.c. \ ,
\end{eqnarray}
where, $y_{\alpha \beta}$  is a symmetry mixing matrix,
 $\overline{\widetilde{L_{L}^{C}}} =\overline{L_{L}^{C}}i\sigma \equiv
 (-\overline{e^{C}},\overline{\nu^{C}} , -\overline{N^{C}},
\overline{E^{C}})$,
 being
\begin{eqnarray}{\label{eq:rotationfordualfield}}
\sigma = T_{2L} + T_{14L} =\dfrac{1}{2} \begin{pmatrix}
0 & -i & 0 & 0 \\ 
i & 0 & 0 & 0 \\ 
0 & 0 & 0 & -i \\ 
0 & 0 & i & 0
\end{pmatrix}  .
\end{eqnarray} 
The scalar decuplet contains  ten degrees of freedom,
using a canonical kinetic term; those can be parametrized as: 
\begin{eqnarray}
\Delta=\begin{pmatrix}
\Delta_{11}^{+} & \Delta_{12}^{++} & \Delta_{13}^{++} & \Delta_{14}^{+} \\ 
\Delta_{21}^{0} & \Delta_{22}^{+} & \Delta_{23}^{+} & \Delta_{24}^{0} \\ 
\Delta_{31}^{0} & \Delta_{32}^{+} & \Delta_{33}^{+} & \Delta_{34}^{0} \\
\Delta_{41}^{+} & \Delta_{42}^{++} & \Delta_{43}^{++} & \Delta_{44}^{+} \\ 
\end{pmatrix} \equiv 
\begin{pmatrix}
\frac{1}{\sqrt{2}}H_{1}^{+} & H_{1}^{++} & \frac{1}{\sqrt{2}}H_{2}^{++} & \frac{1}{\sqrt{2}}H_{3}^{+} \\ 
H_{1}^{0} & -\frac{1}{\sqrt{2}}H_{1}^{+} & -\frac{1}{\sqrt{2}}H_{2}^{+} & \frac{1}{\sqrt{2}}H_{3}^{0} \\ 
-\frac{1}{\sqrt{2}}H_{3}^{0} & \frac{1}{\sqrt{2}}H_{3}^{+} & -\frac{1}{\sqrt{2}}\omega^{+} & -\kappa^{0} \\
\frac{1}{\sqrt{2}}H_{2}^{+} & -\frac{1}{\sqrt{2}}H_{2}^{++} & \rho^{++} & \frac{1}{\sqrt{2}}\omega^{+} \\ 
\end{pmatrix} \ .
\end{eqnarray}
After  EWSB,  the neutral components of the decuplet
develop a VEV and the lagrangian in Eq.~(\ref{eq:Type_II_lag_mass}) becomes:
\begin{eqnarray}{\label{eq:Type_II_lag_mass_after_EWSB}}
 -\mathcal{L}_{\rm{yuk}} &=& y_{\alpha \beta} \bigg ( \overline{\nu_{L \alpha}^{c}} \langle H_{1}^{0} \rangle \
  \nu_{L \beta}  + \dfrac{1}{\sqrt{2}}\overline{\nu_{L \alpha}^{c}}
  \langle H_{3}^{0} \rangle N_{L \beta}  \nonumber \\
&+& \
   \dfrac{1}{\sqrt{2}}\overline{N_{L \alpha}^{c}} \langle H_{3}^{0} \rangle \nu_{L \beta} + \
   \overline{N_{L \alpha}^{c}} \langle \kappa^{0} \rangle N_{L \beta} \bigg ) + h.c \nonumber \\
   &=&  \begin{pmatrix}
   \overline{\nu_{L \alpha}^{c}} & \overline{N_{L \alpha}^{c}} 
   \end{pmatrix} \mathcal{M}
\begin{pmatrix}   
   \nu_{L \beta} \\
   N_{L \beta} 
\end{pmatrix}~, 
\end{eqnarray}
with 
\begin{eqnarray}{\label{eq:Type_II_mass_matrix_after_EWSB}}
\mathcal{M} = \begin{pmatrix}
    y_{\alpha \beta}\langle H_{1}^{0} \rangle  & \frac{1}{\sqrt{2}}y_{\alpha \beta}\langle H_{3}^{0} \rangle \\ 
   \frac{1}{\sqrt{2}}y_{\alpha \beta}\langle H_{3}^{0} \rangle & y_{\alpha \beta}\langle \kappa^{0} \rangle
   \end{pmatrix} \ , 
\end{eqnarray}
with $\alpha$ and  $\beta$  being lepton generation indices.  We demand $ \langle H_3^{0} \rangle < 1$~\gev,
in order to avoid $e-E$ large mixing. The scalar
decuplet will modified the tree-level $\rho$ parameter{~\cite{Langacker:1980js}}.
\begin{eqnarray}{\label{eq:rho_parameter}}
\rho^{\rm{tree}} \simeq 1 - \dfrac{2 \langle H_{1}^{0} \rangle ^{2}}{v^{2} +
  {v^{\prime}}^{ 2} + \langle H_{1}^{0} \rangle^{2} }~.  
\end{eqnarray}
Since, $\rho_{\rm{exp}}=1.00040 \pm 0.00024${~\cite{Agashe:2014kda}}, 
in order to satisfy the $\rho$ constraint, $\langle H_{1}^{0} \rangle \leq
1.5$ GeV.  Notice that $\langle \kappa^{0} \rangle $ is not constrained by $ \rho$.
Assuming $\lbrace \langle H_{1}^{0} \rangle , \langle H_{3}^{0}
\rangle \rbrace <  \langle \kappa^{0} \rangle $, the neutrino
masses for the lightest SM neutrinos  and the heavy ones at second
order in perturbative diagonalization takes the form:
\begin{eqnarray}{\label{eq:Neutrino_masses_interacting_basis_1}}
\mathcal{M}_{\rm{Light}} &=&  y_{\alpha \beta}\langle H_{1}^{0} \rangle
-\dfrac{\langle H_{3}^{0} \rangle^{2}}{\langle \kappa^{0} \rangle
}y_{\alpha \beta}^{-1} y_{\alpha \beta} y_{\alpha \beta}^{\dagger} \ , \\
{\label{eq:Neutrino_masses_interacting_basis_2}}
\mathcal{M}_{\rm{Heavy}} &=&  y_{\alpha \beta}\langle \kappa^{0} \rangle
+\dfrac{\langle H_{3}^{0} \rangle^{2}}{\langle \kappa^{0} \rangle
   }y_{\alpha \beta}^{-1} y_{\alpha \beta} y_{\alpha \beta}^{\dagger} \ .
\end{eqnarray}
In the limit  $  \langle H_{3}^{0} \rangle \ll \langle \kappa^{0}
\rangle $,
the Eq.~(\ref{eq:Neutrino_masses_interacting_basis_1}) and
Eq.~(\ref{eq:Neutrino_masses_interacting_basis_2}) are diagonalized by
the same $U_{\rm{PMNS}}$ mixing matrix~\footnote{The same conclusion is draw for 
Eq.~(\ref{eq:Neutrino_masses_interacting_basis_1}) and
Eq.~(\ref{eq:Neutrino_masses_interacting_basis_2})
forcing $y_{\alpha \beta}$ to be real, assumption 
which is not general, and only will be valid for a real $U_{\rm{PMNS}}$. }.
\begin{eqnarray}{\label{eq:Neutrino_masses_physical_basis_1}}
\mathcal{M}_{\rm{\nu}}^{\rm{diag}} &=& U_{PMNS}^{\dagger}\mathcal{M}_{\rm{Light}}U_{PMNS} =  \langle H_{1}^{0}
\rangle \ U_{PMNS}^{\dagger} \ \mathbf{Y} \ U_{PMNS} \ , 
\\
{\label{eq:Neutrino_masses_physical_basis_2}}
\mathcal{M}_{\rm{N}}^{\rm{diag}} &=& U_{PMNS}^{\dagger}\mathcal{M}_{\rm{Heavy}}U_{PMNS} =  \langle \kappa^{0}
\rangle \ U_{PMNS}^{\dagger} \ \mathbf{Y} \ U_{PMNS} \ ,
\end{eqnarray}
where leptonic indices has been suppressed in matrix $\mathbf{Y}$.
Since both matrices; $\mathcal{M}_{\rm{Heavy}}$ and $\mathcal{M}_{\rm{Light}}$ are 
diagonalized by the same matrix, then 
the heavy neutral leptons (exotics)
and the lightest (SM ones)
has the same mass hierarchy.
Therefore,
\begin{eqnarray}{\label{eq:rel_heavy_lightes}}
\mathcal{M}_{\rm{Ni}}^{\rm{diag}} &=& \dfrac{ \langle \kappa^{0}
\rangle}{\langle H_{1}^{0} \rangle} \mathcal{M}_{\rm{\nu}i}^{\rm{diag}} \ ,
\nonumber \\
\begin{pmatrix}
m_{N1} & 0 & 0 \\
0 & m_{N2} & 0 \\
0 & 0 & m_{N3} 
\end{pmatrix} &=& \dfrac{ \langle \kappa^{0}
\rangle}{\langle H_{1}^{0} \rangle} \begin{pmatrix}
m_{\nu1} & 0 & 0 \\
0 & m_{\nu2} & 0 \\
0 & 0 & m_{\nu3} 
\end{pmatrix} \ .
\end{eqnarray}
Using the data from neutrino
oscillation{~\cite{Aguilar-Arevalo:2013pmq}},
the  lightest of the sterile neutrino satisfies $M_{N1}> 1$ eV.
From this we derived the next constraints on the VEV
of the scalar decuplet.
\begin{eqnarray}{\label{eq:Limits_VEV}}
\langle \kappa^{0} \rangle > \dfrac{1 \ \rm{eV} \langle H_{1}^{0} \rangle}{m_{\nu1}}
\end{eqnarray}
Assuming for instance $m_{\nu1} \simeq \sqrt{\Delta m_{12}^2} \simeq 8.717
\times 10^{-3}$~eV, which is the maximum possible value for $m_{\nu1}$ 
in the normal hierarchy (NH) scenario{\cite{Forero:2014bxa}}, then $\langle \kappa^{0}
\rangle > \  114.707 \ \langle H_{1}^{0} \rangle $,
is a lower bound on $\langle \kappa^{0} \rangle$ 
derived from neutrino physics.
In this model, LFV processes  such as $\mu^{-} \to
e^{+}e^{-}e^{-}$,
$\tau^{-} \to e^{+}e^{-}e^{-}$, $\tau^{-} \to \mu^{+}\mu^{-}\mu^{-}$ 
are mediated by $H_{1}^{++}$ at the tree-level. These processes are controlled by  $y_{\alpha \beta}$ and 
also depends of the new scalar sector spectrum.
\begin{eqnarray}{\label{eq:Br_mu_eee}}
\rm{BR}(\mu^{-} \to  e^{+}e^{-}e^{-}) &\approx& \dfrac{\Gamma(\mu^{-} \to
  e^{+}e^{-}e^{-})}{\Gamma(\mu^{-} \to
  e^{+}\nu_{\mu}\overline{\nu_{e}} )} \ ,  \nonumber \\
&=& \dfrac{1}{(M_{H_{1}^{++}})^{4} G_{F}^{2}} |y_{\mu e}|^{2} |y_{e e}|^{2}   \ .
\end{eqnarray}
\begin{figure}[t!]
\centering
\includegraphics[scale=0.5]{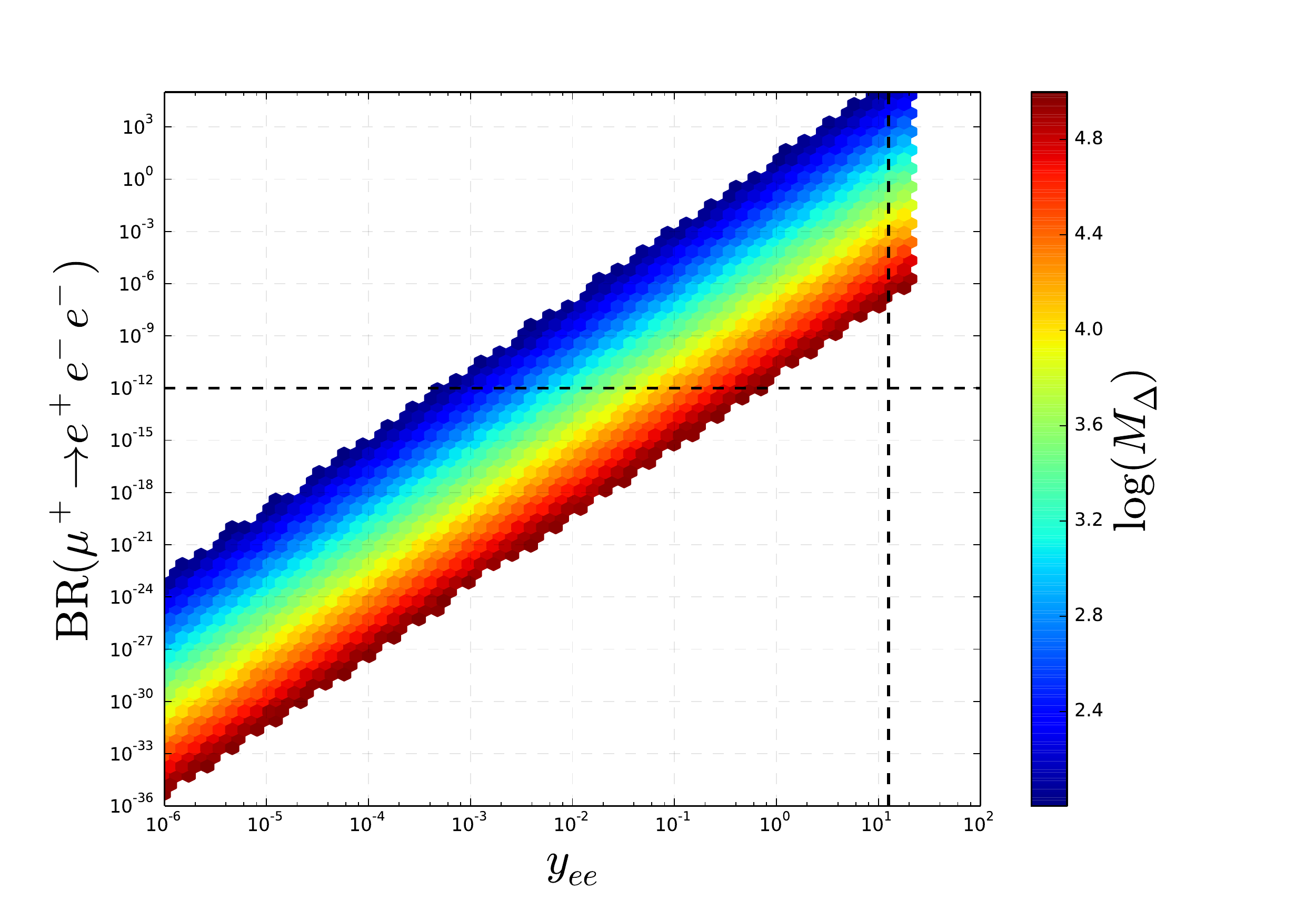}
\caption{$\rm{BR}(\mu^{-} \to e^{+} e^{-} e^{-})$ as a function of
  $y_{ee}$. The vertical dashed line represent the point where
  couplings of order $\sim 4 \pi$ are expected, and the horizontal
    dashed line is the upper limit  for 
    $\rm{BR}(\mu^{-} \to e^{+} e^{-} e^{-})$ process. } 
{\label{fig:Brvsyee}}
\end{figure}
 $\rm{BR}(\mu^{-} \to e^{+} e^{-} e^{-})$ is
  constrained{~\cite{Bellgardt:1987du}} to satisfy  $\rm{BR}(\mu^{-}
  \to e^{+} e^{-} e^{-})< 1.0 \times 10^{-12}$, which is the most
  severe limit.
In figure~\ref{fig:Brvsyee} is displayed the $\rm{BR}(\mu^{-} \to
e^{+} e^{-} e^{-})$ as a function of $y_{ee}$.
The vertical dashed line are the points with yukawa couplings
of order $\sim 4\pi$, which represents the perturbative 
limit. To the left of that line  neutrino
masses and mixing are explained. The points 
with  $y_{ee}> 4 \pi $  are ruled out by perturbativity.
The horizontal dashed line represents the upper 
limit on $\rm{BR}(\mu^{-} \to e^{+} e^{-} e^{-})$, above 
that limit the points are ruled out. 
All the points in the plot were obtained performing a scan of the following
parameters in the range
\begin{eqnarray}
100 \ \rm{GeV} < &m_{H_{1}^{++}} &< 100 \ \rm{TeV} \ , \nonumber \\
10^{-9} \ \rm{GeV} < &\langle H_{1}^{0} \rangle& < 1.5 \ \rm{GeV} \ , \nonumber \\
10^{-9} \ \rm{GeV} < &\langle H_{3}^{0} \rangle& < 1 \ \rm{GeV} \ , \nonumber \\  
10^{-7}  < &y_{\alpha \beta}& < 2 \times 10^{1} \ .  \nonumber 
\end{eqnarray}
All the points in figure~\ref{fig:Brvsyee} satisfy
 the neutrino mixing and masses constraints{~\cite{Forero:2014bxa}} at 
$2\sigma$\footnote{ we only consider the case for the NH scenario.}.
  On the other hand, notice that model $F$ account
   for neutrino masses and mixing extending it with two fermion $15-$plet
   per lepton generation. Since the fermion 15-plet mixes
with the charged 4-plet leptons, then  tree-level LFV processes mediated
by the neutral gauge bosons ( Z, $Z^{\prime}$ and $Z^{\prime \prime}$)
are present. The model will also have restrictions coming from colliders constraints 
on heavy exotic leptons. This model is very interesting, its phenomenology 
is more richer than the two other realizations shown before, but is beyond
scope this work and will be considered in a future work.

\section{Conclusions}\label{sec:conclusions}

In this paper, we provide a mechanism to explain the origin of neutrino
 masses and mixing
in the $SU(4)_{L} \otimes U(1)_{X}$ electroweak extension of the SM through
the tree level realization of the Weinberg-like operator.
For the
model $F$, we construct two of these realization and show how the
masses are generated. For the canonical seesaw, even when
the model predicts the existence of a mixing between the
SM neutrinos and the heavy ones, those sector are
uncorrelated due to the absence of significantly data 
on the neutrino sector. Implying that
heavy neutrinos, if they exist, must be heavy.
For the type II like seesaw model, the introduced decuplet
account for the mixing and masses of the SM neutrinos and
predicts that the exotic neutral leptons has the same mass
hierarchy and mixing pattern than the lightest neutrinos.
In this scenario, the neutral components of the scalar
decuplet (except $\langle \kappa^{0} \rangle $) potentially modified
the $e-E$ mixing and the tree level $\rho$ parameter.
The lower experimental limit established for
the mass of the lightest exotic neutrino give us a lower bound
on $\langle \kappa^{0} \rangle $. It is worth to mention
that the  study done in this paper does not take into account
the analysis of the full scalar sector of the model, mainly because
of the complexity of the scalar potential.
The model give rises to tree level LFV processes, being $\mu \to 3e$
the most sensitive, which is mediated by $H_{1}^{++}$.
Since we do not evaluate neither the full scalar potential nor
the scalar spectrum, then, there are not considerations regarding collider
signatures. However, the model posses signatures worth of exploring.
In the best case scenario ($m_{H_{1}^{++}}$ being the lightest of the
exotic scalars  and small mixing in the full scalar potential)
the signal  $p \ p \to H_{1}^{++} H_{1}^{--} \to  l^{+} l^{+}  l^{-} l^{-}$
will be the promising channel{~\cite{Han:2015hba}} to find the $H_{1}^{++}$.
The phenomenology done through this
paper for model $F$, shall be analogous to the model $E$. For model $A$,
when the tree level realization is the canonical seesaw mechanism, requires
the introduction of three right handed neutrinos ( 15-plet fermions) per lepton generation,
the model will contain a total of 18 neutral leptons, 3 of them being
the SM neutrinos and 15 of them being exotic,
Neutrino masses and mixing are explained within  this model.

The model $B$ is trivial, in the sense that after
extending it  with a fermion singlet, scalar decuplet and a fermion 15-plet
we reach the same phenomenology of the SM when the type-I, the type-II and the type-III
seesaw mechanism are considered.

\end{document}